\documentclass[useAMS,usenatbib]{mn2e}
\usepackage{mathptmx}
\usepackage{dcolumn}
\usepackage{graphicx,color,psfrag}
\usepackage{multirow, threeparttable}
\usepackage{subfig}

\newcolumntype{d}{D{.}{.}{-1}}

\graphicspath{{.}}
\newcounter{todo}
\renewcommand\thetodo{\Alph{todo}}
\def\todo#1{\addtocounter{todo}{1}[[\thetodo: #1]]\strut\vadjust{%
\kern-\dp\strutbox{\vtop to \dp\strutbox{\baselineskip\dp\strutbox\vss\rlap{%
\hskip\hsize\ \rm{$\leftarrow$\thetodo}}\null}}}}
\def\note#1{\strut\vadjust{\kern-\dp\strutbox{\vtop to \dp\strutbox{%
\baselineskip\dp\strutbox\vss\rlap{\hskip\hsize\ {\tiny\rm #1}}\null}}}}
\title[AMI observations of unmatched \emph{Planck} ERCSC LFI sources at 15.75\,GHz]{AMI
observations of unmatched \emph{Planck} ERCSC LFI sources at 15.75\,GHz\thanks{We request that any reference to this paper cites ``AMI Consortium: Perrott et al. 2011"}}

\label{firstpage}

 \author[Perrott et~al.]{AMI Consortium: Yvette~C.~Perrott$^{1}\thanks{Issuing author: email -- ycp21@mrao.cam.ac.uk}$, David~A.~Green$^{1}$, Matthew~L.~Davies$^{1}$, \newauthor 
 Thomas~M.~O.~Franzen$^{2}$, Keith~J.~B.~Grainge$^{1,3}$, Michael~P.~Hobson$^{1}$, \newauthor
 Natasha~Hurley-Walker$^{4}$, Anthony~N.~Lasenby$^{1,3}$, Malak~Olamaie$^{1}$, Guy~G.~Pooley$^{1}$, \newauthor
 Carmen~Rodr\'{i}guez-Gonz\'{a}lvez$^{1}$, Richard~D.~E.~Saunders$^{1,3}$, Anna~M.~M.~Scaife$^{5,6}$, \newauthor 
 Michel~P.~Schammel$^{1}$, Paul~F.~Scott$^{1}$, Timothy~W.~Shimwell$^{1}$,  David~J.~Titterington$^{1}$, \newauthor
 Elizabeth~M.~Waldram$^{1}$\\ 
 $^1$ Astrophysics Group, Cavendish Laboratory,
      19 J.~J.~Thomson Avenue, Cambridge CB3 0HE \\
 $^2$ CSIRO Astronomy \& Space Science, Australia Telescope National Facility, PO Box 76, Epping, NSW 1710, Australia \\
 $^3$ Kavli Institute for Cosmology Cambridge,
      Madingley Road, Cambridge CB3 0HA\\
 $^4$ International Centre for Radio Astronomy Research, Curtin Institute of Radio Astronomy, 1 Turner Avenue, Technology Park, Bentley, WA 6845, Australia \\
 $^5$ Dublin Institute for Advanced Studies, 31 Fitzwilliam Place,
      Dublin 2, Ireland \\
 $^6$ School of Physics and Astronomy, University of Southampton, Highfield, Southampton, SO17 1BJ
 \\
}

\date{Accepted ---; received ---; in original form \today}

\pagerange{\pageref{firstpage}--\pageref{lastpage}}

\pubyear{2011}

\begin{document}

\maketitle

\begin{abstract}
The \emph{Planck} Early Release Compact Source Catalogue includes 26 sources with no obvious matches in other radio catalogues (of predominantly extragalactic sources).  Here we present observations made with the Arcminute Microkelvin Imager Small Array (AMI SA) at 15.75\,GHz of the eight of the unmatched sources at $\delta > +10^{\circ}$.  Of the eight, four are detected and are associated with known objects.  The other four are not detected with the AMI SA, and are thought to be spurious.

\end{abstract}


\section{Introduction}

The \emph{Planck} Early Release Compact Source Catalogue (ERCSC) consists of compact sources detected at each of the \emph{Planck} frequency bands, covering a range from 30 to 857\,GHz.  The Low Frequency Instrument (LFI; 30-, 44- and 70-\,GHz frequency bands) sources are matched against archival data at lower frequencies both for validation purposes and the construction of spectral energy distributions (\citealt{2011arXiv1101.2041P}).  26 sources in the \emph{Planck} ERCSC are reported as having no plausible match in existing, lower frequency radio catalogues of primarily extragalactic sources.  \citet{2011arXiv1101.1721P} conclude that the majority of the unmatched sources are either spurious, explained by extended Galactic structures or have very unusual spectra.  

The Arcminute Microkelvin Imager Small Array (AMI SA) is a radio interferometer specifically designed to have good sensitivity to low-surface-brightness, extended features.  It operates at 15.75\,GHz, relatively close in frequency to the \emph{Planck} LFI.  We therefore decided to observe the eight of the unmatched sources which are visible to the AMI SA, having J2000 $\delta > +10^{\circ}$ (see Table~\ref{tab:srclist}), consisting of sources detected at 44 or 70\,GHz (none of the unidentified sources detected at 30\,GHz were accessible).  The full-width at half-maximum (FWHM) of the AMI SA primary beam is $\simeq 20$\,arcmin so its field of view is comparable to the \emph{Planck} beam sizes of $\simeq$27 and $\simeq$13\,arcmin at 44 and 70\,GHz respectively; the AMI SA is also sensitive to angular scales up to $\simeq 10$\,arcmin so will be able to detect extended objects visible to \emph{Planck} that may have been resolved out in some surveys.

\section{Observations and Data Reduction}

The AMI SA is situated at the Mullard Radio Astronomy Observatory, Cambridge (\citealt{2008MNRAS.391.1545Z}). It consists of ten 3.7-m-diameter dishes with a baseline range of $\simeq 5$--20\,m and observes in the band 12--18\,GHz with eight 0.75-GHz bandwidth channels.  In practice, the lowest two frequency channels are unused due to a low response in this frequency range, and interference from geostationary satellites.  The FWHM of the SA synthesised beam (for combined channel maps) is $\simeq 2$\,arcmin; figures shown include the synthesised beam, which is an effective measure of the resolution.

The eight sources listed in Table~\ref{tab:srclist} were observed from 2011 Mar 21 -- 23.  Each source was observed twice, for one hour at a time, at different hour angles in order to improve the $uv$-coverage.  PLCKERC044 G105.43$-$07.07 and PLCKERC044 G181.40$+$56.02 were also reobserved for 6 and 16 hours respectively on 2011 Aug 13 -- 14 in order to improve extent and spectral index constraints (see Section~\ref{S:constraints}).

\begin{table}
\centering
\caption{The \emph{Planck} unidentified sources observed with the AMI SA, and the sources used as phase calibrators.  \emph{Planck} source names contain the frequency of detection (i.e. PLCKERC044 indicates a source detected at 44\,GHz) and the galactic coordinates of the source.}\label{tab:srclist}
\setlength{\tabcolsep}{4pt}
\begin{tabular}{lccc}\hline
\emph{Planck} ID  & RA  & $\delta$ & Phase  \\
           & (J2000) & (J2000) & calibrator\\  \hline
PLCKERC044 G075.52$-$08.69 & 20 55 58.6 & +31 46 11 & J2109+3532\\
PLCKERC044 G105.35$+$09.85 & 21 42 49.2 & +66 03 11 & J2125+6423\\
PLCKERC044 G105.43$-$07.07 & 22 55 58.8 & +51 49 43 & J2322+5057\\
PLCKERC044 G181.40$+$56.02 & 10 18 32.5 & +39 39 39 & J1023+3948\\
\hline
PLCKERC070 G053.99$-$10.45 & 20 08 54.6 & +13 24 23 & J2016+1632 \\
PLCKERC070 G095.46$+$45.89 & 15 39 06.7 & +61 17 10 & J1551+5806\\ 
PLCKERC070 G120.02$+$09.88 & 00 13 08.2 & +72 32 19 & J0019+7327\\
PLCKERC070 G158.33$-$20.53 & 03 29 03.6 & +31 17 21 & J0329+2756\\ \hline
\end{tabular}
\end{table}


Data reduction was performed using the local software tool \textsc{reduce}, which flags interference, shadowing and hardware errors, applies phase and amplitude calibrations and Fourier transforms the correlator data to synthesize the frequency channels, before output to disk in $uv$ \textsc{fits} format.  Flux calibration was performed using short observations of 3C48, 3C286 or 3C147 near the beginning and end of each run.  The assumed flux densities for 3C286 were converted from Very Large Array total-intensity measurements provided by R. Perley (private communication), and are consistent with the \citet{1987Icar...71..159R} model of Mars transferred on to absolute scale, using results from the \emph{Wilkinson Microwave Anisotropy Probe}.  The assumed flux densities for 3C48 and 3C147 are based on long-term monitoring with the AMI SA using 3C286 for flux calibration (see Table~\ref{tab:Fluxes-of-3C286}).  A correction for changing airmass is also applied using a noise-injection system, the `rain gauge'.

\begin{table}
\centering
\caption{Assumed I~+~Q flux densities of 3C286, 3C48 and 3C147.} \label{tab:Fluxes-of-3C286}
\begin{tabular}{ccccc}\hline
 Channel & $\bar{\nu}$/GHz & $S^{\rm 3C286}$/Jy & $S^{\rm 3C48}$/Jy & $S^{\rm 3C147}$/Jy \phantom{$S^{\rm{X^{X}}}$} \\ \hline
 3 & 13.88 & 3.74 & 1.89 & 2.72 \\
 4 & 14.63 & 3.60 & 1.78 & 2.58 \\
 5 & 15.38 & 3.47 & 1.68 & 2.45 \\
 6 & 16.13 & 3.35 & 1.60 & 2.34 \\
 7 & 16.88 & 3.24 & 1.52 & 2.23 \\
 8 & 17.63 & 3.14 & 1.45 & 2.13 \\ \hline
\end{tabular}
\end{table}

Bright, nearby point sources selected from the Very Long Baseline Array Calibrator Survey (\citealt{1998ASPC..144..155P}) were observed during each observation at hourly intervals for phase calibration purposes (see Table~\ref{tab:srclist} for phase calibrators used for the AMI SA observations).  The reduced visibility data were imaged using \textsc{aips}\footnote{\texttt{http://aips.nrao.edu/}}, from combined channel datasets (for channels 3 to 8 inclusive), with a central frequency of 15.75\,GHz.  Errors on AMI SA flux density values were estimated by adding in quadrature the r.m.s. map noise ($\sigma_{\rm{rms}}$, measured from the \textsc{clean}ed maps), and the error on flux calibration (including rain-gauge correction) of $\simeq 5$ per cent.

%

\section{Results and Discussion}

Maps displayed are not corrected for attenuation due to the primary beam; the flux densities reported have been so corrected.  Where spectral indices $\alpha$ are quoted, the convention $S \propto \nu^{-\alpha}$ is used, where $S$ is flux density and $\nu$ is frequency.  Errors quoted are 1$\sigma$; $\sigma_{\rm{Planck}}$ refers to the positional error appropriate to each \emph{Planck} source, as quoted in the ERCSC.

Data from the following surveys have been used for reference: the VLA Low-Frequency Sky Survey (VLSS, \citealt{2007AJ....134.1245C}), the 7C survey (7C, \citealt{1995A&AS..110..419V}), the Westerbork Northern Sky Survey (WENSS, \citealt{1997A&AS..124..259R}), the NRAO VLA Sky Survey (NVSS, \citealt{1998AJ....115.1693C}), the Green Bank 4.85\,GHz survey (GB6, \citealt{1996ApJS..103..427G}) and the Cosmic Lens All-Sky Survey (CLASS, \citealt{2003MNRAS.341....1M}).  The \textsc{specfind v2.0} catalogue was also used (\citealt{2010A&A...511A..53V}).

\subsection{Sources detected and associations with known objects}
Four of the sources are detected by the AMI SA and are associated with known optical sources, all of which are in the New General Catalogue (NGC, \citealt{1888MmRAS..49....1D}).   These are described below and listed in Table~\ref{T:detections}.

\begin{figure}
  \begin{center}
    \includegraphics[bb=10 47 532 534, clip=, width=0.9\linewidth]{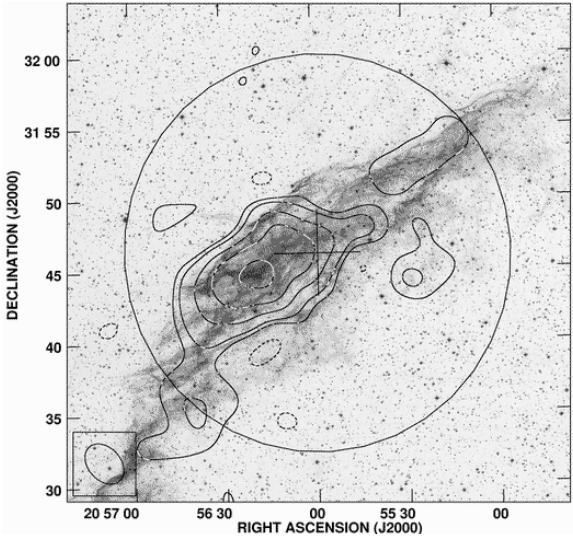}
    \caption{\textbf{PLCKERC044 G075.52$-$08.69:} AMI SA data is shown as contours over DSS greyscale of the north-eastern part of the Cygnus Loop supernova remant.  Contours are at $-$5, 5, 10, 25, 50, 100 $\sigma_{\rm{rms}}$ (see Table~\ref{T:detections}).  The position of the \emph{Planck} detection is marked with a cross, the FWHM of the \emph{Planck} beam is shown as the large ellipse and the AMI SA synthesised beam is shown in the bottom left corner.}
    \label{Fi:cyg_loop}
  \end{center}
\end{figure}

\noindent{\bf PLCKERC044 G075.52$-$08.69}  An extended source is detected at 121$\sigma_{\rm{rms}}$ (see Table~\ref{T:detections}) in the AMI SA map, with the peak of the emission at 2.6$\sigma_{\rm{Planck}}$ from the \emph{Planck} position.  This is associated with NGC\,6992, the north-eastern portion of the Cygnus Loop supernova remnant (see, for example, \citealt{1990AJ....100.1927G}).  Fig.~\ref{Fi:cyg_loop} shows Digital Sky Survey (DSS) red data overlaid with AMI SA contours.  Due to the extent of the source, the AMI SA flux densities are complicated by flux loss, and making a meaningful comparison between the AMI SA and \emph{Planck} data is beyond the scope of this paper.

\begin{figure}
  \begin{center}
    \includegraphics[bb=10 47 531 539, clip=, width=0.9\linewidth]{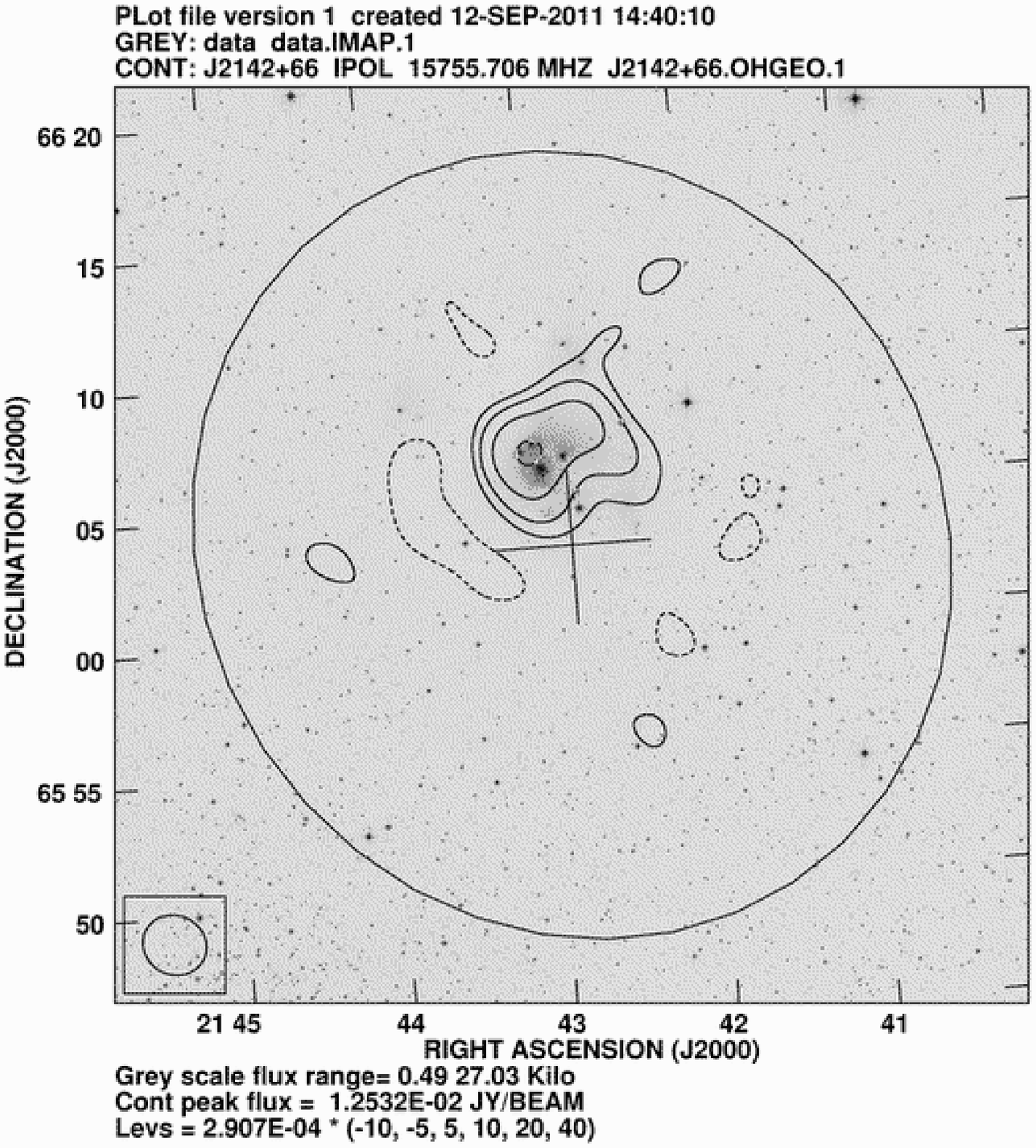}
    \caption{\textbf{PLCKERC044 G105.35$+$09.85:} AMI SA data is shown as contours over DSS greyscale of the reflection nebula NGC\,7133.  Contours are at $-$5, 5, 10, 20, 40 $\sigma_{\rm{rms}}$ (see Table~\ref{T:detections}).  Annotations are as in Fig.~\ref{Fi:cyg_loop}.}
    \label{Fi:NGC7133}
  \end{center}
\end{figure}

\noindent{\bf PLCKERC044 G105.35$+$09.85}  An extended source is detected at 43$\sigma_{\rm{rms}}$ (see Table~\ref{T:detections}) in the AMI SA map, with the peak of the emission at 2.3$\sigma_{\rm{Planck}}$from the \emph{Planck} position.  This is associated with the reflection nebula NGC\,7133.  Fig.~\ref{Fi:NGC7133} shows DSS red data overlaid with AMI SA contours.  Reflection nebulae are expected to emit at 15.75\,GHz and more strongly at 44\,GHz due to a combination of free-free, vibrational dust and possibly spinning dust emission (\citealt{1998ApJ...494L..19D}; \citealt{2011MNRAS.411.1137C}).  The AMI SA and \emph{Planck} measurements are therefore qualitatively consistent in that the spectrum is rising (see Table~\ref{T:detections}), however a quantitative analysis would be complicated due to flux loss and is beyond the scope of this paper.

\begin{figure}
  \begin{center}
    \includegraphics[bb=10 61 532 553, clip=, width=0.9\linewidth]{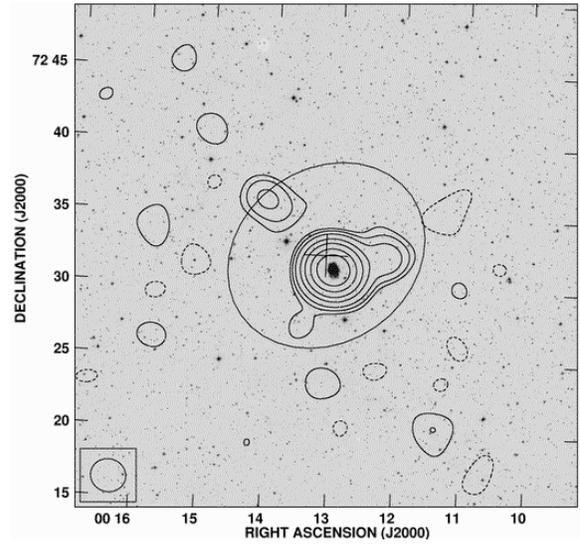}
    \caption{\textbf{PLCKERC070 G120.02$+$09.88:} AMI SA data is shown as contours over DSS greyscale of the planetary nebula NGC\,40.  Contours are at $-$5, 5, 10, 20, 50, 100, 200, 350 $\sigma_{\rm{rms}}$ (see Table~\ref{T:detections}).  Annotations are as in Fig.~\ref{Fi:cyg_loop}.}
    \label{Fi:NGC40}
  \end{center}
\end{figure}

\noindent{\bf PLCKERC070 G120.02$+$09.88}  A point source is detected at 647$\sigma_{\rm{rms}}$ (see Table~\ref{T:detections}) in the AMI SA map with the peak of the emission at 1.4$\sigma_{\rm{Planck}}$from the \emph{Planck} position.  This is associated with the planetary nebula NGC\,40.  Fig.~\ref{Fi:NGC40} shows DSS red data overlaid with AMI SA contours.  NGC\,40 is a well-studied planetary nebula (see, for example, \citealt{2011MNRAS.411.1395L}; \citealt{2011ApJ...738..174M}), which was first identified by Herschel in 1788 (\citealt{1789RSPT...79..212H}).  Although it is compact to the AMI SA beam, its position on the rim of the supernova remnant CTA\,1 (e.g. see \citealt{2011arXiv1108.4156S}) makes constructing a radio spectrum for it difficult because of the extra extended emission which may be measured by single dishes, depending on the background subtraction method used; for a recently compiled spectrum at radio frequencies below 44\,GHz see \citeauthor{2011arXiv1108.4156S}.  A search of the \emph{Planck} catalogue shows probable associations at 100 and 143\,GHz also, at 2.2 and 1.3$\sigma_{\rm{Planck}}$ from the AMI SA position respectively.  The measurement at 143\,GHz (553$\pm$31\,mJy) is lower than the 70\,GHz (688$\pm$135\,mJy) and 100\,GHz (776$\pm$58\,mJy) measurements -- this may be an indication of spinning dust emission, but more observations between 100 -- 1000\,GHz are needed to confirm this.

\begin{figure}
  \begin{center}
    \includegraphics[bb=10 48 533 539, clip=, width=0.9\linewidth]{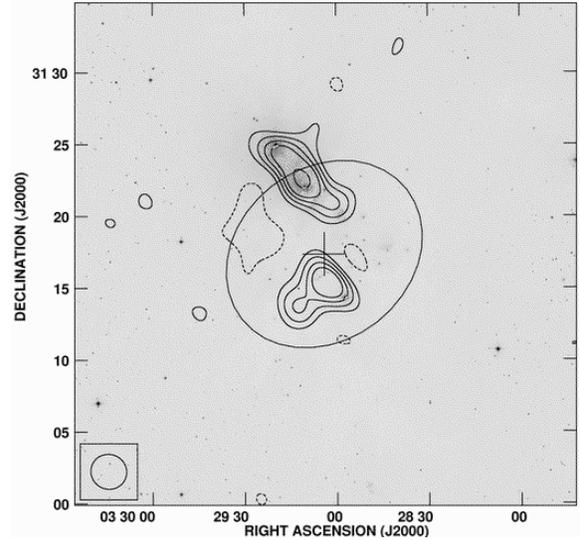}
    \caption{\textbf{PLCKERC070 G158.33$-$20.53:} AMI SA data is shown as contours over DSS greyscale of the reflection nebula NGC\,1333.  Contours are at $-$3, 3, 5, 7, 9, 13 $\sigma_{\rm{rms}}$ (see Table~\ref{T:detections}).  Annotations are as in Fig.~\ref{Fi:cyg_loop}.}
    \label{Fi:NGC1333}
  \end{center}
\end{figure}

\noindent{\bf PLCKERC070 G158.33$-$20.53}  Two lobes of extended emission are detected at 15 and 13$\sigma_{\rm{rms}}$ (see Table~\ref{T:detections}) in the AMI SA map with the peaks of the emission at 5 and 2$\sigma_{\rm{Planck}}$from the \emph{Planck} position respectively.  These are associated with the reflection nebula NGC\,1333.  Fig.~\ref{Fi:NGC1333} shows DSS red data overlaid with AMI SA contours.  As noted above with regard to NGC\,7133, the AMI SA and \emph{Planck} measurements are qualitatively consistent in that the spectrum is rising (see Table~\ref{T:detections}), however quantitative analysis has not been attempted since this is complicated by flux loss.

\begin{table*}
\caption{\emph{Planck} unidentified LFI sources which are detected with the AMI SA, and are associated with known optical sources.  These are supernova remnants (SNR), reflection nebulae (RNe) or planetary nebulae (PNe).  In each case, the distance(s) to the peak(s) of the emission detected with the AMI SA and the peak flux densit(ies) are given.}\label{T:detections}
\centering
\begin{tabular}{lccccccc}
\hline
 \emph{Planck} ID  & \emph{Planck} flux  & \emph{Planck}    & AMI SA peak  & AMI SA & Distance from  & Optical & Object \\
                   & density             & positional error & flux density & $\sigma_{\rm{rms}}$ & \emph{Planck} position & association & type\\ 
 & (mJy) & (arcmin) & (mJy beam$^{-1}$) & ($\mu$Jy beam$^{-1}$) & (arcmin) & & \\ \hline
PLCKERC044 G075.52$-$08.69 & 1769$\pm$165 & 1.7 & 79$\pm$4 & 560 & 4.5 & NGC\,6992$^{\dag}$ & SNR \\
PLCKERC044 G105.35$+$09.85 & 1342$\pm$117 & 1.7 & 14.1$\pm$0.8 & 290 & 3.9 & NGC\,7133 & RNe \\
PLCKERC070 G120.02$+$09.88 & 688$\pm$135 & 0.87 & 350$\pm$18 & 540 & 1.2 & NGC\,40 & PNe \\
\multirow{2}{*}{PLCKERC070 G158.33$-$20.53} & \multirow{2}{*}{2089$\pm$165} & \multirow{2}{*}{1.0} & 7.5$\pm$0.6 & \multirow{2}{*}{410} & 5.4 & \multirow{2}{*}{NGC\,1333} & \multirow{2}{*}{RNe} \\ 
                         &              &     & 5.4$\pm$0.5 & & 1.9 & & \\\hline
\noalign{$^{\dag}$ Part of the Cygnus Loop -- see, for example \citet{1990AJ....100.1927G}.}
\end{tabular}
\end{table*}

\subsection{Non-detections}

Although some sources were detected near the positions of the remaining four \emph{Planck} unidentified sources, these were classed as non-detections.  Flux detection limits are taken as 5$\sigma_{\rm{rms}}$ as measured from the map, with the exception of PLCKERC070 G095.46$+$45.89 which is dynamic-range-limited due to the presence of a bright source; in this case the flux detection limit is taken as the peak flux density of the brightest non-believable feature.

\noindent{\bf PLCKERC044 G105.43$-$07.07}
Two faint (peak flux density $\leq$1.0\,mJy) point sources are detected in the AMI SA map within 1.9\,arcmin (1.0$\sigma_{\rm{Planck}}$) from the \emph{Planck} position.  These are associated with three NVSS point sources with flux densities of 7.4, 8.8 and 8.7\,mJy, indicating a falling spectrum inconsistent with the \emph{Planck} flux density of 1558$\pm$231\,mJy at 44\,GHz.

\noindent{\bf PLCKERC044 G181.40$+$56.02}
Three faint (peak flux density $\leq$1.9\,mJy) point sources are detected in the AMI SA map within 5.4\,arcmin (3$\sigma_{\rm{Planck}}$) from the \emph{Planck} position.  Matching with NVSS shows that all have falling spectra, so they are unlikely to be associated with the \emph{Planck} source which has a flux density of 1234$\pm$232\,mJy.

\noindent{\bf PLCKERC070 G053.99$-$10.45}
A 17$\pm$1\,mJy point source is detected in the AMI SA map at 0.4\,arcmin (0.5$\sigma_{\rm{Planck}}$) from the \emph{Planck} position; this is also detected as a point source in NVSS with a flux density of 66.3$\pm$2.6\,mJy so it is unlikely to be associated with the \emph{Planck} source which has a flux density of 852$\pm$153\,mJy.

\begin{figure}
  \begin{center}
%
%
\begin{psfrags}%
\psfragscanon%
%
\psfrag{s05}[t][t]{\color[rgb]{0,0,0}\setlength{\tabcolsep}{0pt}\begin{tabular}{c}\Large{Frequency (GHz)}\end{tabular}}%
\psfrag{s06}[b][b]{\color[rgb]{0,0,0}\setlength{\tabcolsep}{0pt}\begin{tabular}{c}\Large{Flux density (Jy)}\end{tabular}}%
%
\psfrag{x01}[t][t]{0}%
\psfrag{x02}[t][t]{0.1}%
\psfrag{x03}[t][t]{0.2}%
\psfrag{x04}[t][t]{0.3}%
\psfrag{x05}[t][t]{0.4}%
\psfrag{x06}[t][t]{0.5}%
\psfrag{x07}[t][t]{0.6}%
\psfrag{x08}[t][t]{0.7}%
\psfrag{x09}[t][t]{0.8}%
\psfrag{x10}[t][t]{0.9}%
\psfrag{x11}[t][t]{1}%
\psfrag{x12}[t][t]{\Large{$10^{-1}$}}%
\psfrag{x13}[t][t]{\Large{$10^{0}$}}%
\psfrag{x14}[t][t]{\Large{$10^{1}$}}%
\psfrag{x15}[t][t]{\Large{$10^{2}$}}%
\psfrag{x16}[t][t]{\Large{$10^{3}$}}%
%
\psfrag{v01}[r][r]{0}%
\psfrag{v02}[r][r]{0.2}%
\psfrag{v03}[r][r]{0.4}%
\psfrag{v04}[r][r]{0.6}%
\psfrag{v05}[r][r]{0.8}%
\psfrag{v06}[r][r]{1}%
\psfrag{v07}[r][r]{\Large{$10^{-1}$}}%
\psfrag{v08}[r][r]{\Large{$10^{0}$}}%
%
\resizebox{\linewidth}{!}{\includegraphics{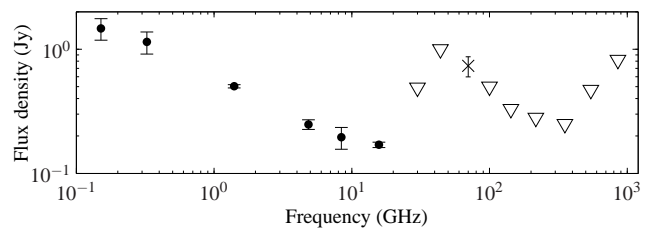}}%
\end{psfrags}%
%

    \caption{The filled circles with error bars show the spectrum of the point source, 6C\,153854$+$612327, detected in the PLCKERC070 G095.46$+$45.89 field.  The cross with error bar shows the \emph{Planck} measurement for comparison purposes, and the downward-pointing triangles are ERCSC flux detection limits at the other \emph{Planck} frequencies.  The \emph{Planck} unidentified source is unlikely to be associated with the source detected by the AMI SA.}
    \label{Fi:J1539+6117_spec}
  \end{center}
\end{figure}

\noindent{\bf PLCKERC070 G095.46$+$45.89}
A 170$\pm$8\,mJy point source is detected in the AMI SA map at 5.9\,arcmin (6.8$\sigma_{\rm{Planck}}$) from the \emph{Planck} position.  This is a well-characterised radio source, 6C\,153854$+$612327, with $\alpha \simeq 0.5$ from 74 to 8400\,MHz (from the \textsc{specfind} catalogue; data points collated from, in order of increasing frequency, VLSS, 7C, WENSS, NVSS, GB6 and CLASS).  Fig.~\ref{Fi:J1539+6117_spec} illustrates the spectrum; the AMI SA flux density agrees well with the lower frequency data points, however the \emph{Planck} flux, included for comparison, of 734$\pm$136\,mJy does not.  If there were a sharp turnover in the spectrum around the \emph{Planck} frequencies, the \emph{Planck} ERCSC flux detection limits (shown as open triangles in Fig.~\ref{Fi:J1539+6117_spec}) show that the source should be detected at frequencies $>$70\,GHz, but there are no other sources within 45\,arcmin in the ERCSC.  Given the distance from the \emph{Planck} position, the well-characterised falling spectrum and the lack of detection at higher frequencies, 6C\,153854$+$612327 is not likely to be associated with the \emph{Planck} source.

\subsubsection{Limits on spectral indices and extent}\label{S:constraints}

\begin{table*}
\caption{\emph{Planck} unidentified LFI sources which are not detected with the AMI SA, and limiting spectral index values for a source the size of the \emph{Planck} beam.  The AMI SA flux detection limits are 5$\sigma_{\rm{rms}}$, except for PLCKERC070 G095.46$+$45.89 which has the flux density of the brightest non-believable feature as its detection limit.  Flux densities are in mJy and sizes are in arcmin.  See text for an explanation of the reliability values and CMBSUBTRACT (CS) flag.}\label{T:non-detections}
\centering
\begin{tabular}{lccccccc}
\hline
 \emph{Planck} ID  & \emph{Planck}  & \emph{Planck}    & \emph{Planck} flux & \emph{Planck} beam & AMI SA flux  & AMI SA flux & Limiting \\
            & reliability & CS & density     & major axis  & detection limit & detection limit & spectral \\ 
            & & flag & (mJy) & (arcmin) & (no $uv$-taper, mJy) & ($uv$-taper = 300$\lambda$, mJy) & index \\ \hline
PLCKERC044 G105.43$-$07.07 & 0.79 & 2 & 1558$\pm$231 & 28.4 & 0.91 & 2.0 & $-$0.73 \\
PLCKERC044 G181.40$+$56.02 & 0.74 & 2 & 1234$\pm$232 & 28.3 & 0.56 & 1.7 & $-$0.37 \\
PLCKERC070 G053.99$-$10.45 & 0.83 & 2 & 851$\pm$153 & 14.7 & 4.7 & 11 & $-$0.88 \\
PLCKERC070 G095.46$+$45.89 & 0.79 & 1 & 734$\pm$136 & 14.7 & 6.2 & 9.1 & $-$0.91 \\ \hline
\end{tabular}
\end{table*}

There are three possible reasons for the non-detection of these unidentified sources, if they are real.  For a compact (to the SA beam), non-variable source, non-detection could be due to a rising spectrum between 15.75\,GHz and the \emph{Planck} frequency of detection.  A limiting spectral index for this case can be calculated using the AMI SA flux detection limit.  For the 44\,GHz sources, the calculated spectral indices are found to be non-physical ($<-$7).  For the 70\,GHz sources they are $<-$3 which would require extreme thermal dust emission, but the flux densities extrapolated to the higher \emph{Planck} frequencies using the spectral index limits are well above the \emph{Planck} flux detection limits at the corresponding frequencies, and these sources are not detected at any other frequencies.  The mean SA beam FWHM taken from the observations of the four non-detected sources is $\simeq$165\,arcsec, so it can be concluded that non-variable sources on this scale corresponding to the \emph{Planck} detections do not exist.

Variability is significant at AMI frequencies, and even more so at \emph{Planck} frequencies (see, for example, Fig.~11 of \citealt{2011arXiv1101.1721P}).  Many \emph{Planck} sources are expected to be blazars, and detections near the flux density limits are more likely to be in a flare state.  \citet{2011ApJS..194...29R} monitored the fluxes of 1158 blazars with the Owens Valley Radio Observatory (OVRO) 40\,m telescope at 15\,GHz, on a bi-weekly basis for over three years.  The largest peak-to-trough amplitude ratio found in the sample was $\simeq$43, with a time scale of just under a year between the peak and trough.  The AMI SA observations were taken between just under one to two years after the \emph{Planck} observations, so the non-detected \emph{Planck} sources could potentially be blazars detected in a flare state by \emph{Planck}, while the AMI SA observations were made during a low state.

In the case of the 44\,GHz sources, multiplying the AMI SA flux detection limits by 43 still implies extreme spectral indices between 15.75 and 44\,GHz of $<-$3.6 and $<-$3.8 respectively.  Two of the three 44\,GHz receivers are located at the opposite side of the \emph{Planck} focal plane from the third, which is next to the 30 and 70\,GHz receivers, and observe the same point on the sky roughly a week apart in time from the other receivers.  It is therefore conceivable that a blazar in a flare state could be observed at 44\,GHz but not at the neighbouring frequencies.  However, in that case the effect of averaging between the three 44\,GHz receivers could only lower the reported flux, requiring the spectral indices between 15.75 and 44\,GHz to be even more extreme.  In the case of the 70\,GHz sources, multiplying the AMI SA flux detection limits by 43 implies spectral indices between 15.75 and 70\,GHz of $<-$0.96 and $<-$0.68 respectively.  The 100\,GHz receivers are located next to the 70\,GHz receivers, and the \emph{Planck} flux detection limit at 100\,GHz is 344\,mJy, so in order for these sources not to be detected simultaneously at 100\,GHz the spectra must turn over to have indices $>$2 between 70 and 100\,GHz.  Similar spectra are observed (see Fig.~5 of \citealt{2011arXiv1101.1721P}) so this possibility cannot be ruled out.  It should be noted, however, that blazars with a high variability amplitude are found to be more likely to be gamma-ray-loud (\citealt{2011ApJS..194...29R}), and none of these sources are matched in the Fermi-Large-Area-Telescope first source catalogue (\citealt{2010ApJS..188..405A}).

Alternatively, if a source is extended on scales larger than the SA beam, flux will be ``resolved out'' due to under-sampling of the large spatial scales.  All of the sources are indicated in the \emph{Planck} ERCSC as being compact to the \emph{Planck} beam and should therefore be of the size of the beam or smaller.  An upper limit on the spectral index for each source can be calculated for which a Gaussian source of the same size as the \emph{Planck} beam would be undetected in the AMI SA map.  For simplicity, the limit is calculated using a circular Gaussian source with FWHM equal to the \emph{Planck} major axis beam size quoted in the catalogue.

To calculate the flux loss for a Gaussian source with a given FWHM, a simulated source was sampled with the real $uv$-coverage from the observations then mapped in \textsc{aips} to recover a peak flux density.  To recover more flux on extended scales, a ``$uv$-taper'' of 300$\lambda$ was applied.  This is a Gaussian weighting function which down-weights the longer baselines, with a distance of 300$\lambda$ to the 30\% point, where $\lambda$ is the central wavelength of the observation.  The peak flux density recovered can be compared with the peak flux required for detection on the $uv$-tapered AMI SA maps.  Flux limits and limiting spectral indices are quoted for each source in Table~\ref{T:non-detections}.  The spectral index limits are not implausible, however the spectra implied are nearly all very steeply rising, and are upper limits since if the source is smaller than the beam size the ``true'' spectral index limits will be even more negative.

Each source in the ERCSC is assigned a ``reliability'' value between 0 and 1, corresponding to the probability that a source lying in a patch of sky with a given sky noise will have an estimated flux density accurate to within 30\% (\citealt{2011arXiv1101.2041P}).  The reliability values for these undetected sources (see Table~\ref{T:non-detections}) are comparable to the lowest reliability value in the entire catalogue, 0.74.  Additionally, all of these sources have a ``CMBSUBTRACT'' flag set to either 2 or 1 in the ERCSC.  This means that they are either not detected in the CMB subtracted maps, or are detected with a flux density difference $>$30\% compared to the flux density detected on the non-subtracted maps, respectively.  In contrast, the unidentified sources that were detected with the AMI SA have reliability values ranging from 0.84 -- 0.96, and only two have ``CMBSUBTRACT'' flags set to 1, while the other two are set to 0.  Thus although the AMI SA observations cannot rule out the possibility that the non-detections are either variable or real, extended sources with rising spectra, it seems more likely that these unidentified sources are spurious.

\section{Conclusions}
Of the eight unidentified \emph{Planck} LFI sources observable with the AMI SA, four are detected at 15.75\,GHz and are associated with known objects.  The other four are not detected.  Calculations of limiting spectral indices show that the undetected sources cannot be non-variable and compact at or below the scale-size of the AMI SA beam.  The possibility that they are either variable or extended sources with steeply rising spectra between 15.75\,GHz and 44 or 70\,GHz cannot be ruled out by the AMI SA observations, however combined with the relatively low \emph{Planck} reliability values and ``CMBSUBTRACT'' flags for these sources, it seems likely that they are spurious.

\section*{Acknowledgments}
We thank the staff of the Mullard Radio Astronomy Observatory for their invaluable assistance in the commissioning and operation of AMI, which is supported by Cambridge University and the STFC.  We also thank the anonymous referee for useful comments on this paper.  The Digitized Sky Survey was produced at the Space Telescope Science Institute under U.S. Government grant NAG W-2166. The images of these surveys are based on photographic data obtained using the Oschin Schmidt Telescope on Palomar Mountain and the UK Schmidt Telescope. The plates were processed into the present compressed digital form with the permission of these institutions.  This research has made use of the SIMBAD database, operated at CDS, Strasbourg, France.  YCP acknowledges the support of a CCT/Cavendish Laboratory studentship.  MLD, TMOF, CRG, MO, MPS and TWS acknowledge the support of STFC studentships.

\bsp \label{lastpage}

\end{document}